%% file: epl.tex
\documentstyle[epsf]{europhys}
\input euromacr

\input mssymb
\newcounter{saveeqn}
\newcommand{\alpheqn}{\setcounter{saveeqn}{\value{equation}}%
\setcounter{equation}{0}%
\addtocounter{saveeqn}{1}%
\renewcommand{\theequation}{\mbox{\arabic{saveeqn}\alph{equation}}}%
}
\newcommand{\reseteqn}{\setcounter{equation}{\value{saveeqn}}%
\renewcommand{\theequation}{\arabic{equation}}}

\begin{document}

\newcommand{\D}{\displaystyle}
\newcommand{\SA}{\scriptstyle}
\newcommand{\SB}{\scriptscriptstyle}
\newcommand{\T}{\textstyle}
\newcommand{\NM}{\negmedspace}
\newcommand{\NT}{\negthickspace}
\euro{43}{2}{207-212}{1998}
\Date{}
\shorttitle{Finite temperature  $t\!-\!J$ model in large $d$}
\title{Finite temperature hole dynamics in the $\bf t\!-\!J$ model: exact
results for high dimensions}
\author{D.E. Logan \And M.P.H. Stumpf}
\institute{University of Oxford, Physical and Theoretical Chemistry
Laboratory\\ South Parks Road, Oxford OX1 3QZ, UK}
\rec{6 April 1998}{5 June 1998}
\pacs{
\Pacs{71}{10.Fd}{Lattice fermion models (Hubbard model, etc.)}
}
\maketitle
\begin{abstract}
We discuss the dynamics of a single hole in the $t\!-\!J$ model at
finite temperature, in the limit of 
large spatial dimensions. The problem is shown to yield a simple and
physically transparent solution, that exemplifies the continuous
thermal evolution of the underlying string picture from the $T=0$
string-pinned limit through to the paramagnetic phase.
\end{abstract}

Following early pioneering work [1,2], and given added impetus by
the discovery of high-$T_{c}$ superconductivity, the study of
single-particle excitations in magnetic insulators remains highly
topical. At possibly its simplest, the dynamics of a single hole in an
antiferromagnet (AF) is captured by the $t - J$ model [3], itself the
strong  coupling limit of the Hubbard model and a paradigm for the
physics of strong, local electron correlations: hole motion, with
nearest neighbour (NN) hopping amplitude $t$, occurs in a restricted
subspace of no doubly occupied sites, and in a background of AF
coupled spins with NN exchange couplings $J=4t^{2} /U$. The string
picture (see {\it e.g.} [4,5]), wherein the hole is pinned by the string of
upturned spins its motion creates, has proven central in understanding
the $t - J$ model at $T=0$: since spins in the resultant string are
flipped relative to the N\'eel configuration, resulting for $J>0$ in
energetically unfavourable exchange fields that generate a potential
growing linearly with distance from its point of creation, the hole is
thereby confined. And while string unwinding, and hence a finite hole
mobility, is in general induced both by spin-flip interactions and
Trugman loop motion [6], the underlying picture is robust even in
$d=2$ dimensions (see {\it e.g.}[5]). 

  The limit of large spatial dimensions, $d= \infty$, has also proven
highly fruitful in study of the $t\! -\! J$ model [7-11]. Here the string picture
has been shown to be exact, and the model solved at $T=0$ [7,8]; the
additional roles of disorder [8], and of second NN hopping [9], have
also been investigated. Moreover, in addition to providing exact
solutions, the $d^{ \infty}$ limit serves as a starting point for
$1/d$ expansions, enabling a systematic approach to the finite-$d$ case.

  Given the success and utility of the large-$d$ approach, we use it here to
investigate another key element of hole dynamics in the t-J model: the
role of temperature. That the spectrum of single-particle excitations,
and hole conductivity, should change radically with $T$ on the scale
of the N\'eel temperature $T_{N}$, is easily understood; for while the
two-sublattice A/B structure characteristic of the N\'eel ordered AF
phase persists for $0\! <\! T\! <\! T_{N}$, the sublattice magnetization is not
saturated: $m(T)\! <\! m(0)\! =\! 1/2$. In contrast to the $T\!=\!0$ limit, therefore,
\it any \rm site has a non-zero probability $p_{ \alpha \sigma}$
(where $\alpha $=A or B distinguishes the sublattices) of being
occupied by an electron of either spin type, $\sigma = +/-$ or 
$\uparrow\!/\! \downarrow$. In
consequence, single-step hole motion may incur either an exchange 
energy penalty --- if a spin hops to the `wrong' sublattice, as occurs
inevitably at $T\!=\!0$ --- or an energy {\it gain} if the converse
occurs; and since spins in the string need not therefore be
energetically penalized, a finite hole mobility results. Hole dynamics
thus change character entirely on the scale of $T_{N}$ itself, since
in the paramagnetic phase for $T\!>\! T_{N}$, where $m(T)=0$, the energy
barrier to hole motion vanishes, and the hole propagates essentially as a free
particle. In this Letter we show that an exact, yet rather
simple and physically transparent, solution to the problem can be
given; providing in addition a benchmark against which approximations,
such as the NCA [10,11], can be judged.

The $t\! -\! J$ Hamiltonian $\hat{H}\! =\! \hat{H}_t\! +\! \hat{H}_J$ is given in
standard notation by
\begin{equation}
\hat{H} = -t\sum_{i,j,\sigma}
\tilde{c}^\dagger_{i\sigma}\tilde{c}_{j\sigma}
+ {\T \frac{1}{2}} J \sum_{i,j} {\bf{\hat{S}}}_i\!\cdot\!
{\bf{\hat{S}}}_j\ \ \  ; 
\end{equation}
the sums are over NN sites, and
$\tilde{c}^\dagger_{i\sigma}\!=\!c^\dagger_{i\sigma}(1\!-\!\hat{n}_{i
-\sigma})$ embodies the constraint of no double occupancy by
electrons. To ensure a well behaved limit for $d\!=\!\infty$, the hopping
and exchange couplings are scaled as
\begin{equation}
t=t_*/\sqrt{Z}\ \ , \ \ J=J_*/Z
\end{equation}
with $Z\!\rightarrow\! \infty$ the coordination number ($Z\!=\!2d$ for a
hypercubic lattice). Spin-flip interactions are wholly suppressed for
$d^\infty$ (they contribute to $O(1/d)$), see {\it eg.}[7,8]; whence $\hat{H}_J
\! \!\rightarrow\!\! \frac{1}{2} J \sum_{i,j}\hat{S}_i^z \hat{S}_j^z$, simple
molecular field theory (MFT) for which becomes trivially exact
for $d^\infty$. To study hole dynamics we consider the thermally
averaged, local one-hole Green function (retarded), given by
\begin{equation}
\overline{G}_{ii}(\omega) = \sum_s
P(s)\langle i;s|(z-\hat{H}')^{-1} |i;s\rangle \ \ ;
\end{equation}
$|i;s\rangle$ denotes the hole on site $i$ with configuration
$s\!=\!\{\sigma_k \}$ for the remaining spins, with energy $E_s$ under
$\hat{H}_J$; $\hat{H}'\!\! =\!\! \hat{H}\!\! -\!\! E_s$ and $z\! =\!
\omega\! +\!
i0+$. $G_{ii}(\omega)\! =\! \langle i;s|(z-\hat{H}')^{-1}|i;s\rangle$ is
the local Green function for the given configuration $s$, the
probability $P(s)$ of generating which is given exactly for $d^\infty$
by $P(s)\!=\!\prod_{k(\neq i)} p_{\SB k}(\sigma_{\SB k})$ (with corrections $O(1/d)$ ),
reflecting the statistical independence inherent in $d^\infty$. Here
$p_{\scriptscriptstyle k}(\sigma_{\SB k})$, such that $p_{\SB k}(\sigma)\! =\! p_{\alpha\sigma}$ for site $k$
belonging to the $\alpha\! =\! A$ or $B$ sublattice (with $p_{\SB A
\sigma}\! =\! p_{\SB B -\sigma}$), is given by
\begin{equation}
p_{\alpha \sigma} = {\T \frac{1}{2}} \left[ 1+2 \lambda_\alpha \sigma
m(T)\right]
\end{equation}
with $\lambda_{\alpha}\!=\!+1(-1)$ for $\alpha\!=\!A(B)$; and the sublattice
magnetization is given from simple MFT by
\begin{equation}
m(T)={\T \frac{1}{2}} \tanh\left[\frac{2T_N}{T} m(T)\right]
\end{equation}
with $T_N\!=\!\frac{1}{4}J_*\ (k_B\!=\!1)$ the associated N\'{e}el temperature.

To obtain $G_{ii}(\omega)$ for an arbitrary configuration $s$, as now
sketched, we first introduce a Feenberg self-energy $S_i(\omega)$ via
\begin{equation}
G_{ii}(\omega) = [z-S_i(\omega)]^{-1}\ \ .
\end{equation}
To obtain $S_i(\omega)$ one may proceed thus: (a) Separate the
resolvent operator $\hat{G}\!=\!(z-\hat{H}')^{-1}$ as $\hat{G}\! =\! \hat{G}_0
+ \hat{G}_0\hat{H}_t\hat{G}$, where $\hat{G}_0\!=\!(z-{\hat{H}'}_J)^{-1}$
with ${\hat{H}'}_J\!=\!\hat{H}'\!-\!E_s$, such that $\langle
j;s'|\hat{G}_0|i;s\rangle\! =\! \delta_{ij}\delta_{ss'}/z$; (b) take matrix
elements of $\hat{G}$ between $\langle j;s'|$ and $|i;s\rangle$,
obtaining thereby an `equation of motion' for $G_{ji}(\omega)\!=\!\langle
j;s'|\hat{G}|i;s\rangle$ which (c) may then be iterated perturbatively
in $t$. This generates the Nagaoka path formalism [1,2], enabling
systematic construction of $S_i(\omega)$ in unrenormalized form,
{\it i.e.} as an explicit function of frequency $\omega$. The Nagaoka path
formalism is however a particular case of the Feenberg perturbation
series (see {\it e.g.}[12]), the general power of which is to express
$S_i(\omega)$ in {\it renormalized} form --- {\it i.e.} as an explicit
functional of the $\{G_{jj}\}$ --- enabling from Eq.(6) a self-consistent
solution for the $\{G_{jj}\}$.

The problem is particularly clear for the $Z\! \rightarrow\! \infty$ Bethe
lattice, on which we focus explicitly (the hypercube can also be
handled, the same physical ideas being involved but the algebra more
complex). Here, $S_i(\omega)$ is given by
\begin{equation}
S_i = \sum_{j,s'} \langle i;s| \hat{H}_t|j;s'\rangle\langle
j;s'|(z-\hat{H}')^{-1}| j;s'\rangle \langle j;s'|\hat{H}_t| i;s\rangle
\ \ ,
\end{equation}
where since hopping is solely NN, the spin configuration $s'$, with
energy $E_{s'}$ under $\hat{H}_J$, differs from $s$ only by a {\it
single} hole/electron transfer; and $\langle j;s'|\hat{H}_t
|i;s\rangle \equiv -t$. From the energy change involved in transferring
a $\sigma_j$-spin electron on site $j$ in configuration $s$ to an
empty site $i$, it follows simply that $E_{s'}\! =\! E_s\! +\! \lambda_j
\sigma_j \omega_p(T)$, reflecting either the energy cost $(\lambda_j
\sigma_j\! =\! +1)$ or gain $(\lambda_j\sigma_j=-1)$ under single-step
hole motion; here 
\begin{equation}
\omega_p(T) = J_* m(T)
\end{equation}
is the magnitude of the energy incurred in hole transfer which,
involving necessarily sites on different sublattices, is equivalent to
the exchange energy cost for an on-site spin-flip from $\sigma\!=\! +(-)$
to $-(+)$ on an $A(B)$ sublattice site. From this it follows
straightforwardly that for $Z\rightarrow\infty$,
$\langle j;s'|(z-\hat{H}')^{-1}|j;s'\rangle\! =\!
G_{jj}(\omega-\lambda_j \sigma_j \omega_p(T))$, whence as sought Eq.(7) 
yields $S_i$ as a functional of the $\{G_{jj}\}$:
\alpheqn
\begin{equation} S_i\! =\! t^2\sum_j
G_{jj}(\omega-\lambda_j \sigma_j \omega_p(T))
\end{equation}
(with sites $j$ NN to $i$). The thermal average of
$S_i$, denoted by $\overline{S}$ (and naturally independent of the
sublattice index $i$), follows directly noting likewise that
$\overline{G}_{ii} \equiv \overline{G}$ is independent of $i$:
\begin{equation}
\overline{S} = t_*^2 \sum_\sigma p_{\alpha \sigma}
\overline{G}(\omega-\lambda_{\alpha} \sigma \omega_p(T))
\end{equation}
\reseteqn
Finally, we recognize that $\overline{G}\!=\!\overline{[z-S_i]^{-1}}$
reduces simply to $\overline{G}\!=\![z-\overline{S}]^{-1}$ for
$Z\!\rightarrow\! \infty$ (the corrections being $O(Z^{-1})$). Hence the
desired result
\begin{equation}
\overline{G}(\omega) = \left[z-t_*^2 \left(p_{\SB B\downarrow}
\overline{G}(\omega-\omega_p(T)) + p_{\SB B\uparrow} \overline{G}( \omega
+\omega_p(T)) \right)\right]^{-1}
\end{equation}
where $p_{B\sigma}(\!=\! p_{A-\sigma})$ and $\omega_p(T)$ are given from
Eqs.(4,8) solely in terms of the sublattice magnetization $m(T)$, whose
$T$ and $J_*$ dependence follows explicitly from the simple MFT
equation, Eq.(5).

Eq.(10) is exact and, with Eqs.(4,5,8), closed. Further, although we
have merely sketched the derivation here its physical content is
quite clear, involving simply the thermal probability ($p_{\SB
B\sigma}$ factors) with which a NN site to the initially created hole
is occupied by a $\sigma$-spin electron, together with the associated
energy cost/gain (the $\omega \mp \omega_p(T)$ shifts) incurred in NN
hole transfer. For $T\! =\! 0$, where $m(T)\! =\! \frac{1}{2}$ and hence $
p_{\SB B \sigma}\! =\! \delta_{\SB \sigma \downarrow}$ and
$\omega_p(0)\! =\! \frac{1}{2}J_*$, Eq.(10) reduces to
$\overline{G}(\omega)=[z-t^2_* \overline{G}(\omega-
\frac{1}{2}J_*)]^{-1}$, the well known exact $d^\infty$ result from
the $T\! =\! 0$ string picture [8] (and for finite-$d$ an approximation often
used in practice [13]), yielding the familiar discrete spectrum
characteristic of the confined/localized hole. In the paramagnetic
phase by contrast, $T\! \ge\! T_N\! =\! \frac{1}{4}J_*$, where
$m(T)\! =\! 0\! =\! \omega_p(T)$ and $p_{\SB \alpha \sigma}\! =\!
\frac{1}{2}\  \forall\ \alpha , 
\sigma$, Eq.(10) reduces to $\overline{G}(\omega) = [z-t_*^2
\overline{G}(\omega)]^{-1}$, producing the semielliptic spectrum of
full width $4t_*$ characteristic of free hole motion in the $d^\infty$
Bethe lattice.

As a typical example, Fig.1 shows for $J_*/t_*=0.4$ the resultant hole
spectrum $D(\omega) = -\pi^{-1}\, {\rm Im}\ \overline{G}(\omega)$ at four
different $T$'s; the spectral sum rule $\int_{-\infty}^{\infty} d\omega
D(\omega)=1$ is always satisfied. From Eq.(10),  the thermal evolution
of $D(\omega)$ is controlled by that of the sublattice magnetization
$m(T)$, Eq.(5). At the lowest $T$ shown, $T=0.3 T_N$, $m(T)$ remains
close to saturation: $m(T)/m(0)\simeq0.997$. Hence $D(\omega)$
resembles closely its discrete $T=0$ counterpart: the peak positions
coincide, with only minor thermal broadening. Increasing $T$ to $0.5
T_N$ produces a modest reduction in $m(T)$ to $\sim 0.96 m(0)$, but an
appreciable effect upon $D(\omega)$, with increased broadening and the
incipient formation of a background continuum. The latter, a precursor
of free-hole dynamics, grows with increasing $T$; see Fig.1(c) for
$T=0.7T_N$ where $m(T)\simeq 0.83 m(0)$. By $T=0.9 T_N$ where
$m(T)\simeq0.53 m(0)$, the semiellipse characteristic of free hole
motion is clearly emerging, Fig.1(d), showing also that the `shortest
strings' --- those to the low-$\omega$ side of $D(\omega)$ --- are the
last to be thermally eroded. Finally, in the paramagnet for $T>T_N =
\frac{1}{4}J_*$, the resultant semielliptic $D(\omega)$ is independent
of $T$ for obvious physical reasons.

Fig.1 shows clearly that while the effective spectral width is on the
scale of the hopping $t_*$, the essential character of the spectrum
evolves on the typically much smaller scale of $T_N=\frac{1}{4} J_*$
($=0.1t_*$ in Fig.1). Qualitatively similar behaviour occurs on
varying $J_*/t_*$; the rule of thumb being that for any fixed $T/T_N$
(and hence constant $m(T)$ and $p_{\SA \alpha \sigma}$, see
Eqs.(4,5)), increasing $J_*$ amplifies the hole transfer shifts in
Eq.(10) (since $\omega_p(T)=J_* m(T)$), thus producing a somewhat
`colder' spectrum.

The behaviour described above is reflected also in the hole
conductivity $\sigma(\omega;T)$, itself of $O(1/d)$ in high dimensions
but with $\sigma(\omega;T)/t^2$ 
(or equivalently ${\rm Tr}\sigma_{\lambda\lambda}(\omega;T)$) of order
unity and a much
studied quantity [7,8]. Due to the absence of vertex corrections for
$d^\infty$, the (real) dynamical conductivity $\sigma(\omega;T)$ is
expressible in terms of the hole spectrum $D(\omega)$; and an
exact expression for it may be deduced along the lines sketched above for
$D(\omega)$. We quote only the result (which is independent of lattice
type), namely
\begin{equation}
\sigma(\omega;T)\propto \frac{(\!1\!\! -\! e^{-\beta \omega}\!)}{\omega}
\frac{\int_{-\infty}^{\infty}\! d\omega_1
e^{-\beta\omega_1}D(\!\omega_1\!)\left[p_{\SB
B\downarrow}D\left(\!\omega_1\!+\!\omega\!-\!\omega_p(\!T) \right)+p_{\SB
B\uparrow}D\left(\!\omega_1\!+\!\omega\!+\!\omega_p(\!T)\right) \right]}{\int_{-\infty}^{\infty}
d\omega_1 e^{-\beta\omega_1} D(\omega_1)}
\end{equation}
where $\beta=1/T$ and extraneous constants have been dropped for
clarity. Temperature enters Eq.(11) in two ways: via (a) explicit
Boltzmann factors and (b) the $T$-dependence of $D(\omega)$,
$p_{\SB B\sigma}$ and $\omega_p(T)$. In previous work [8] the latter have
been neglected entirely and replaced by their $T=0$ counterparts with
$p_{\SB B\sigma} = \delta_{\sigma\downarrow}$ and
$\omega(0)=\frac{1}{2}J_*$, whereupon Eq.(11) reduces to the result of
[8]. This amounts to retaining only the $T=0$ N\'{e}el spin
configuration in calculating $\sigma(\omega;T)$, but is not a
physically realistic approximation: since $D(\omega)$ for $T=0$ is
discrete, the dynamical conductivity consists of a series of
$\delta$-functions whose weight alone depends on T [8], and the dc
conductivity $\sigma_0(T) = \sigma(\!\omega\!=\!0;T)$ vanishes for all
$T$. While the introduction of disorder [8] broadens the $\delta$-peaks
and leads to non-trivial behaviour, the absence of a proper
$T$-dependence limits somewhat the utility of this approximation.

Eq.(11) by contrast is exact at finite-$T$. To illustrate it, Fig.2
gives the resultant dc conductivity $\sigma_0(T)/\sigma_0(T_N)$ {\it vs.}
$T/T_N$ for $J_*/t_*=0.4$. As expected it is $T_N=\frac{1}{4}J_*$, and not
the hopping $t_*$ ($=10 T_N$ in Fig.2), that sets the scale for the
$T$-dependence of the conductivity in the AF phase. $\sigma_0(T)$ is
non-zero for any $T>0$, reflecting the fact that true hole
confinement/localization is exclusively a $T=0$ phenomenon, as
embodied also in the corresponding loss of the discrete $T=0$ hole
spectrum at finite $T$ (Fig.1); the progressive increase of
$\sigma_0(T)$ with $T$ in the AF phase likewise reflects the enhanced
thermal broadening of $D(\omega)$. At $T=T_N$ the conductivity
acquires a sharp cusp, across which $d\sigma_0(T)/dT$ changes sign. In the
paramagnetic phase, $p_{\SB \alpha\sigma}=\frac{1}{2}$, $\omega_p=0$
and (see Eq.(10)) $D(\omega)$ is independent both of $T$ and $J_*$:
Eq.(11) for $\sigma(\omega;T)$ thus reduces in practice to its well
known $J_*=0$ limit [7,8]. For $T>T_N$ the $T$-dependence of
$\sigma_0(T)$ is thus controlled by the {\it sole} remaining scale,
$t_*$, diminishing monotonically with increasing $T$ and approaching
$\sigma_0(T)\sim1/T$ behaviour for $T/t_* \gtrsim 1\! - \! 2$ (which
asymptote is evident from the $\beta=1/T$ prefactor in Eq.(11) for
$\omega=0$).

In summary we have shown that in the large dimensional limit, the
finite temperature dynamics of a single hole in an $AF$, as described
by the $t\!-\!J$ model, admits to a simple closed solution that
captures in a physically transparent fashion the continuous thermal
evolution of the underlying string picture, from the $T=0$
string-pinned limit to the paramagnetic phase $T>T_N$ where there
is no exchange energy barrier to hole propagation. Although for
clarity we have focussed explicitly on the Bethe lattice, the
analysis is readily extended both to lattices with non-retraceable
paths and, relatedly, to the case of a second NN hopping, which even
for a Bethe lattice involves non-retraceable loop paths and likewise
admits to a closed solution for all $T$ [14]. The additional effects of
disorder, as hitherto considered for $T=0$ [8], are also easily
encompassed; this, together with an elaboration of the present work,
will be described in a subsequent paper.
\stars

\newpage
\vspace{2cm}

 \begin{figure}
\leavevmode
\vskip-7.5cm
\epsfxsize=15cm
\epsffile{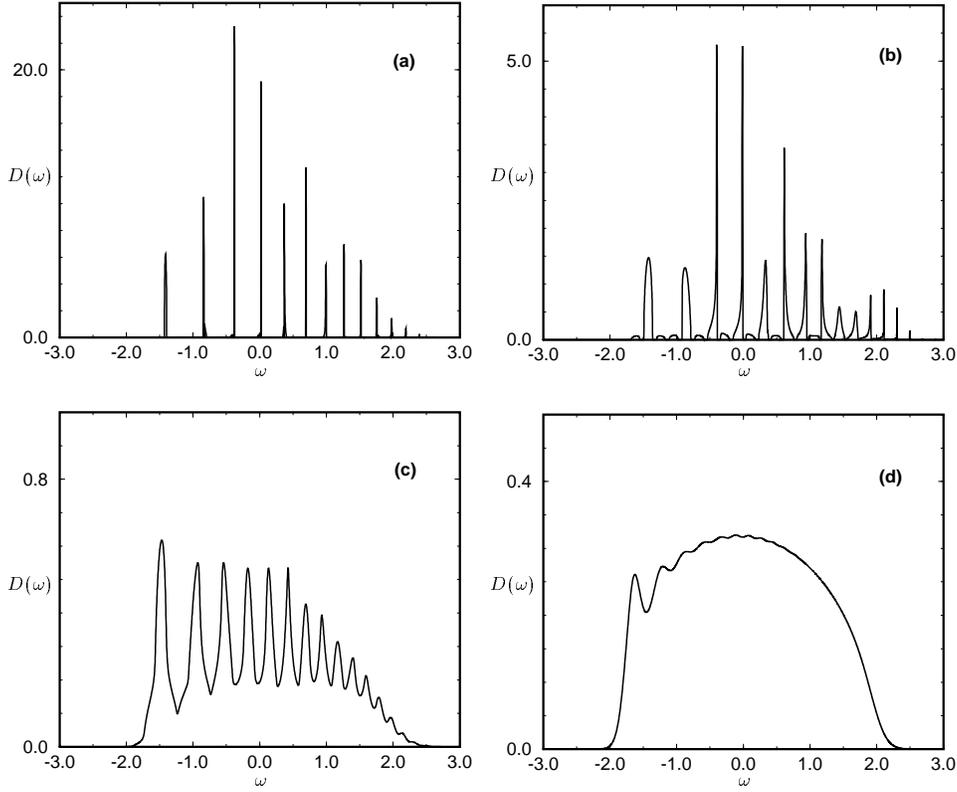}
\vskip2.5cm

\caption{Hole spectrum $D(\omega)$ {\it vs.} $\omega/t_*$ for
$J_*/t_*=0.4$ and $T/T_N=0.3$ (a), $0.5$ (b), $0.7$ (c) and $0.9$ (d);
the N\'{e}el temperature $T_N=\frac{1}{4}J_*$.}
\label{fig1}
\end{figure}
\vspace{-4cm}
\begin{figure}
\vskip-4cm
\epsfxsize=7cm
\epsffile{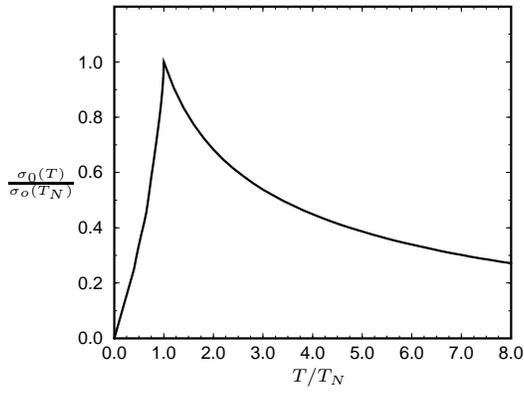}

\vskip-0.8cm
\hskip3.5cm
$\SA T/T_N$

\vskip-3.cm
\hskip-.3cm
$\SA \frac{\sigma_0(T)}{\sigma_o(T_N)}$
%
\vskip3.0cm
\caption{Static conductivity $\sigma_0(T)/\sigma_0(T_N)$ {\it vs.}
$T/T_N$ for $J_*/t_*=0.4$. Note that $T_N=t_*/10$.}
\label{fig2}
\end{figure}

\end{document}

%% file: euromacr.tex
\def\And{{\rm and\ }}

\def\stars{\bigskip\centerline{***}\medskip}

\newif\ifboo \boofalse

\def\Review#1{\boofalse{\it #1},}
\def\Name#1{{\sc #1},}
\def\Vol#1{\ifboo Vol. {\bf #1}\else{\bf #1}\fi}
\def\Year#1{\ifboo #1\else(#1)\fi}
\def\Book#1{\bootrue{\it #1},}
\def\Page#1{\ifboo {\rm p. #1}\else{\rm #1}\fi}